\documentclass[prl, amsmath, amssymb, reprint]{revtex4-1}
\usepackage{graphicx}% Include figure files
\usepackage{dcolumn}% Align table columns on decimal point
\usepackage{bm}% bold math
\usepackage{hyperref}
\usepackage{amsmath} %for matrices
\usepackage{float}
\usepackage[section]{placeins}

\preprint{prl/123-QED}

\begin{document}

\title{Nonlinear Optical Trapping of Gold Nanoparticles}

% \author{S. Mirzaei-Ghormish}
% \affiliation{Department of Electrical and Computer Engineering, Brigham Young University, Provo, UT}

% \author{K. Qaderi}
% \affiliation{Independent Researcher, Pasadena, CA}

% \author{D.Smalley}
% \affiliation{Department of Electrical and Computer Engineering, Brigham Young University, Provo, UT}
\author{S. Mirzaei-Ghormish$^{1}$, K. Qaderi$^{1}$, and D. Smalley$^{1}$}
\email{smalley@byu.edu}
\affiliation{$^1$Department of Electrical and Computer Engineering, Brigham Young University, Provo, UT}

\begin{abstract}

This paper presents the most complete framework to date for understanding the nonlinear optical trapping of highly absorbing nanoparticles within the dipole regime. Highly absorbing and plasmonic particles garner considerable interest due to their enhanced optical interactions, yet traditional linear theory fails to predict their trapping behavior, particularly in the longitudinal direction. We introduce a more comprehensive theory that incorporates four-wave mixing and two-photon absorption, addressing significant gaps in the current literature. Our research shows TPA to be crucial for longitudinal trapping stability. Furthermore, our findings propose a novel phenomenon of multiple split traps in both saturable absorption and reverse saturable absorption regimes. Our framework extends beyond existing models by including factors such as self-induced back-action and scattering forces, aligning theoretical predictions with experimental observations, and laying a robust foundation for future nonlinear optical trapping and manipulation research.

\end{abstract}

\maketitle

\section{Introduction}

In levitated optomechanics, plasmonic particles, particularly gold nanoparticles, are recognized for their capacity to enhance light-matter interactions via localized surface plasmon resonances \cite{mirzaei2023optical}. Contrary to predictions from linear trapping theory (which posits that highly absorbing particles should be unstable at the intensity maxima of a single laser beam due to overpowering repulsive forces) experimental work has shown that gold nanoparticles can indeed be stably trapped at these points \cite{svoboda1994optical,seol2006gold,hansen2005expanding}.  This stability has been attributed to surface creeping waves or non-spherical particle shapes reducing scattering force \cite{furukawa1998optical, brzobohaty2015three}; while useful, these explanations are insufficient for our purposes as micrographs of the particles in experiments we seek to explain how highly spherical morphologies\cite{svoboda1994optical,seol2006gold,hansen2005expanding} and creeping waves do not account for longitudinal stability \cite{furukawa1998optical}. Furthermore, under femtosecond pulse illumination, gold nanoparticles exhibit unique trapping behaviors such as split potential wells in the transverse plane and circumgyration in potentials with radii smaller than the diffraction limit, which defy linear optical trapping theories \cite{jiang2010nonlinear,zhang2018nonlinearity,zhu2023nonlinear, qin2021nonlinearity}. These phenomena, which hint at the involvement of two-photon absorption (TPA), point to the need for a more robust theoretical framework that can account for the full range of experimental observations, including the stable trapping of spherical gold nanoparticles in both the longitudinal and transverse planes.

\begin{figure}
\centering
\includegraphics[width=.40\textwidth]{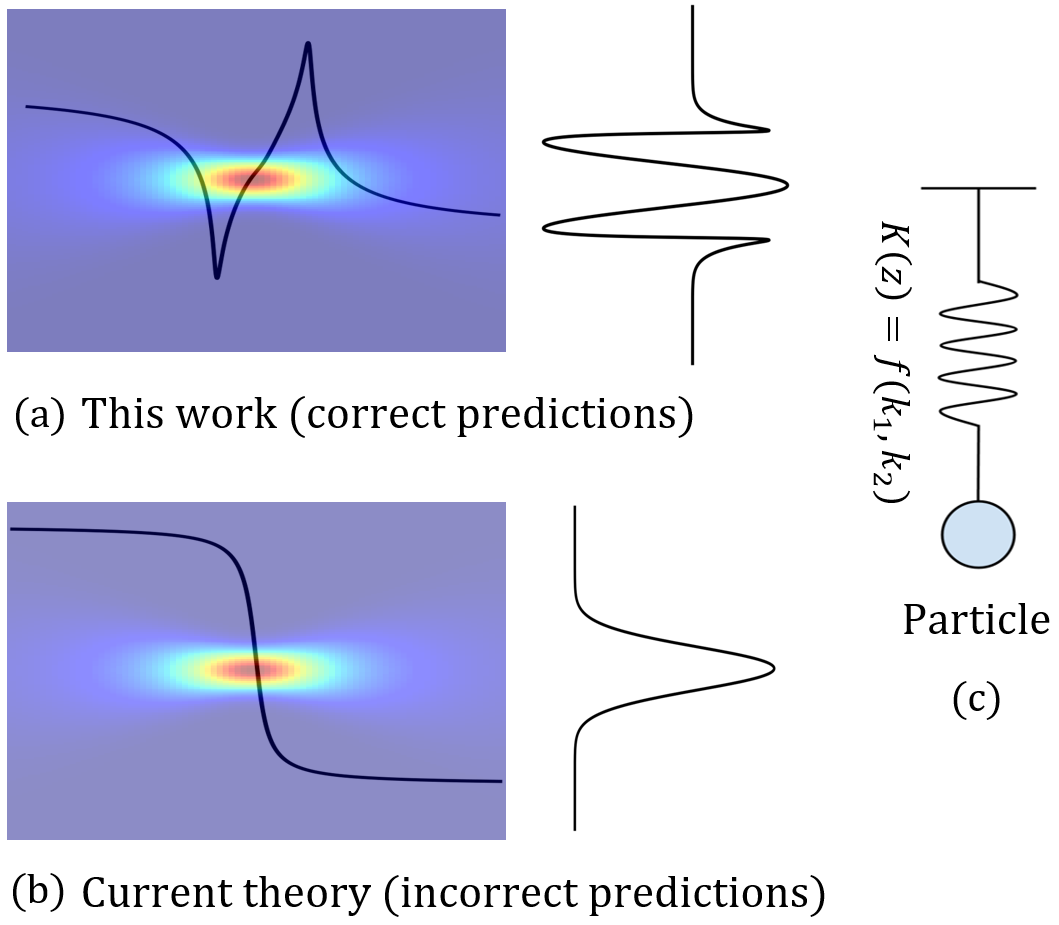}
\caption{The comparison between linear (current theory) and nonlinear (this work) models for potential traps. (a) Shows nonlinear longitudinal and transverse potentials, and (b) shows linear longitudinal and transverse potentials. (c) The nonlinear trapping system behaves like a nonlinear harmonic oscillator, where, the total stiffness $k(z)$ is a function of linear $k_{1}$ and nonlinear $k_{2}$ counterparts.}
\label{comparision}
\end{figure}

In this paper, for the first time, we show that TPA accounts for the stable trapping of gold nanoparticles in the longitudinal direction, as well as the split traps in the transverse plane. To this end, we present a comprehensive theory for nonlinear optical trapping of all types of nanoparticles, including non-absorbing, weakly-absorbing, and highly-absorbing nanoparticles. We investigate the effects of four-wave mixing (FWM) and TPA on optical trapping, separately. Our proposed theory not only aligns with prior experimental findings but also predicts new phenomena. Within the transverse plane, we illustrate how TPA creates a dual-split trap under the saturable absorption (SA) regime and a tri-split trap within the reverse saturable absorption (RSA) regime. Furthermore, we show that, unlike the transverse plane, TPA creates a stable asymmetric longitudinal potential in both SA and RSA regimes, where the stable point occurs in front of a focal point of a tightly focused Gaussian.  Finally, we present the physical interpretation of the nonlinear trap system, likening it to a nonlinear harmonic oscillator and associating its parameters with linear and nonlinear susceptibilities.
The theory findings are summarized in \ref{comparision}.

 Gold nanoparticles exhibit strong third-order nonlinearity at plasmon resonances \cite{de2008optical, wang2010size, elim2006observation, zhang2017nonlinear}. To explore the effects of this nonlinearity on trapping, our analysis assumes a linearly polarized femtosecond laser ($\lambda=532$ nm, repetition rate: $\nu=80$ MHz, pulse duration: $100$ fs) that propagates in the $z$-direction to illuminate gold nanoparticles ($R=20$ nm) immersed in water. Using the dipole approximation, the optical forces for a monotonic wave can be expressed as \cite{novotny2012principles}:

\begin{equation}
\vec{F}(\vec{r}) = \frac{\alpha^{\prime}}{2} \vec {\nabla } \langle |\vec{E}(\vec{r},t)|^2 \rangle + \alpha ^{\prime\prime} \omega \langle \vec{E}(\vec{r},t) \times \vec{B}(\vec{r},t) \rangle
\label{equation1}
\end{equation}

Where $\alpha=\alpha^{\prime}+i\alpha ^{\prime\prime}
= {\alpha_{0}}/(1-i\frac{\alpha_{0}k^3}{6\pi\epsilon_{0}\epsilon_{m}})$ denotes the effective polarizability of the particle. Meanwhile, $\alpha_{0}=4\pi\epsilon_0R^3\frac{\epsilon_{p}-\epsilon_{m}}{\epsilon_{p}+2\epsilon_{m}}$ stands for the static polarizability, with $\epsilon_{p}$ and $\epsilon_{m}$ being the relative permittivities of the particle and medium, respectively. The first term in equation \ref{equation1} corresponds to the gradient force, whereas the second term signifies the scattering force. Under the influence of a  pulsed laser or high-intensity continuous-wave laser, the induced nonlinear polarizability becomes significant and must be taken into account in the computation of optical forces. Here we do not consider the nonlinear effect of the medium as it is much smaller than that of the particle. The relative permittivity of the particle by including the third-order nonlinearity is 
$\epsilon_{p}=1+\chi_{1}+3\chi_{3}E^{2}$, where $\chi_{1}=\chi^{\prime}_{1}+i\chi^{\prime\prime}_{1}$ is the complex linear susceptibility, $\chi_{3}=\chi^{\prime}_{3}+i\chi^{\prime\prime}_{3}$ is the complex nonlinear susceptibility, in which the real part corresponds to FWM, the imaginary part relates to TPA, and $E^2$ is the amplitude of the electric field. The third-order nonlinearity of gold originates from interband transitions, intraband transitions, and hot electrons \cite{boyd2014third}. Both intraband and interband transitions have an instantaneous response, while hot electrons exhibit a response time on the order of hundreds of femtoseconds. 

Details of the calculations are presented in the Supplemental Information. Here, we present the final results. The real and imaginary parts of effective polarizability in terms of both linear and nonlinear static polarizabilities read
\begin{equation}
    \alpha^{\prime} = \alpha_{o,r}^{(L)}+\alpha_{o,r}^{(NL)}
     \label{equation2}
\end{equation}

\begin{equation}
    \alpha^{\prime\prime} = (\alpha_{0,I}^{(L)}+\alpha_{0,I}^{(NL)})+\frac{k^3}{6\pi\epsilon_{o}}\left[ (\alpha_{0,r}^{(L)})^2+(\alpha_{0,I}^{(L)})^2 \right]
     \label{equation3}
\end{equation}
where the real $ \alpha_{0,r}^{(L)}$ and imaginary  $\alpha_{0,I}^{(L)}$ components of linear static polarizability are expressed by

\begin{equation}
    \alpha_{0,r}^{(L)} = 4\pi\epsilon_oR^3\frac{\chi^{\prime}(3+\chi^{\prime})+(\chi^{\prime\prime})^2}{(3+\chi^{\prime})^2+(\chi^{\prime\prime})^2}
    \label{equation4}
\end{equation}

\begin{equation}
    \alpha_{0,I}^{(L)} = 12\pi\epsilon_oR^3\frac{\chi^{\prime\prime}}{(3+\chi^{\prime})^2+(\chi^{\prime\prime})^2}
    \label{equation5}
\end{equation}
and the real $\alpha_{0,r}^{(NL)}$ and imaginary $\alpha_{0,I}^{(NL)}$ components of nonlinear static polarizability  are written as follows: 
\begin{equation}
    \alpha_{0,r}^{(NL)} = \beta[(3+\chi^\prime )^2\chi_3^{\prime} + 2(3+\chi^\prime)\chi^{\prime\prime}\chi_3^{\prime\prime} - \chi_3^{\prime}(\chi^{\prime\prime})^2]
    \label{equation6}
\end{equation}

\begin{equation}
    \alpha_{0,I}^{(NL)} = \beta[(3+\chi^\prime)^2\chi_3^{\prime\prime} - 2(3+\chi^\prime)\chi^{\prime\prime}\chi_3^{\prime} - \chi_3^{\prime\prime}(\chi^{\prime\prime})^2]
    \label{equation7}
\end{equation}
where $\beta=(36\pi\epsilon_oR^3E^2)/[(3+\chi^\prime)^2+(\chi^{\prime\prime})^2]^2.$

 In low absorption limit ($\chi^{\prime\prime} \ll \chi^\prime$, $\chi_3^{\prime\prime}\ll \chi_3^\prime$),  more compact equations can be achieved as follows:
 
\begin{equation}
    \alpha^{\prime} = 4\pi\epsilon_oR^3\frac{\epsilon^{\prime}_{p}-1}{\epsilon^{\prime}_{p}+2} + 36\pi\epsilon_oR^3E^2\chi_3^{\prime}\frac{\epsilon^{\prime}_{p}-1}{(\epsilon^{\prime}_{p}+2)^2}
    \label{equation8}
\end{equation}
\begin{equation}
    \alpha^{\prime\prime} = \alpha_1^{\prime\prime}+12\pi\epsilon_0R^3\frac{\chi^{\prime\prime}+3E^2\chi_3^{\prime\prime}}{(\epsilon^{\prime}_{p}+2)^2}
    \label{equation9}
\end{equation}
where
\begin{equation}
    \alpha_1^{\prime\prime} = \frac{8}{3}\pi\epsilon_ok^3R^6(\frac{\epsilon^{\prime}_{p}-1}{\epsilon^{\prime}_{p}+2})^2 + 48\pi\epsilon_ok^3R^6E^2\chi_3^{\prime}\frac{\epsilon^{\prime}_{p}-1}{(\epsilon^{\prime}_{p}+2)^3}
    \label{equation10}
\end{equation}
As shown, the FWM contributes to both gradient and scattering forces. Conversely, TPA solely contributes to the scattering force. The ratio of the FWM to TPA components in equation \ref{equation9} is proportional to $(kR)^3\left(\frac{\chi^{\prime}_3}{\chi^{\prime\prime}_3}\right)$, where $(kR)^3 \ll 1$ and $\left(\frac{\chi^{\prime}_3}{\chi^{\prime\prime}_3}\right) \gg 1$. As a result, in certain circumstances, the effect of FWM dominates, whereas in others, TPA primarily contributes to the scattering force. Depending on the values of $\chi^{\prime}_{3}$ and $\chi^{\prime\prime}_{3}$, as well as the magnitude of the input power, the FWM and TPA effects can either increase or decrease the gradient and scattering forces. Hence, the new potential traps rather than the harmonic quadratic potential can be achieved. 

From equations \ref{equation8} to \ref{equation10}, both the real and imaginary components of effective polarizability can be formulated by a combination of linear and nonlinear components. Therefore, the gradient and scattering forces can also be depicted as the summation of their linear and nonlinear counterparts. This suggests that we can decompose the system into two coupled linear and nonlinear oscillators. Later, we provide a physical interpretation of the nonlinear oscillator by defining its parameters in terms of linear and nonlinear susceptibility.
 
\begin{figure}
\centering
\includegraphics[width=.48\textwidth]{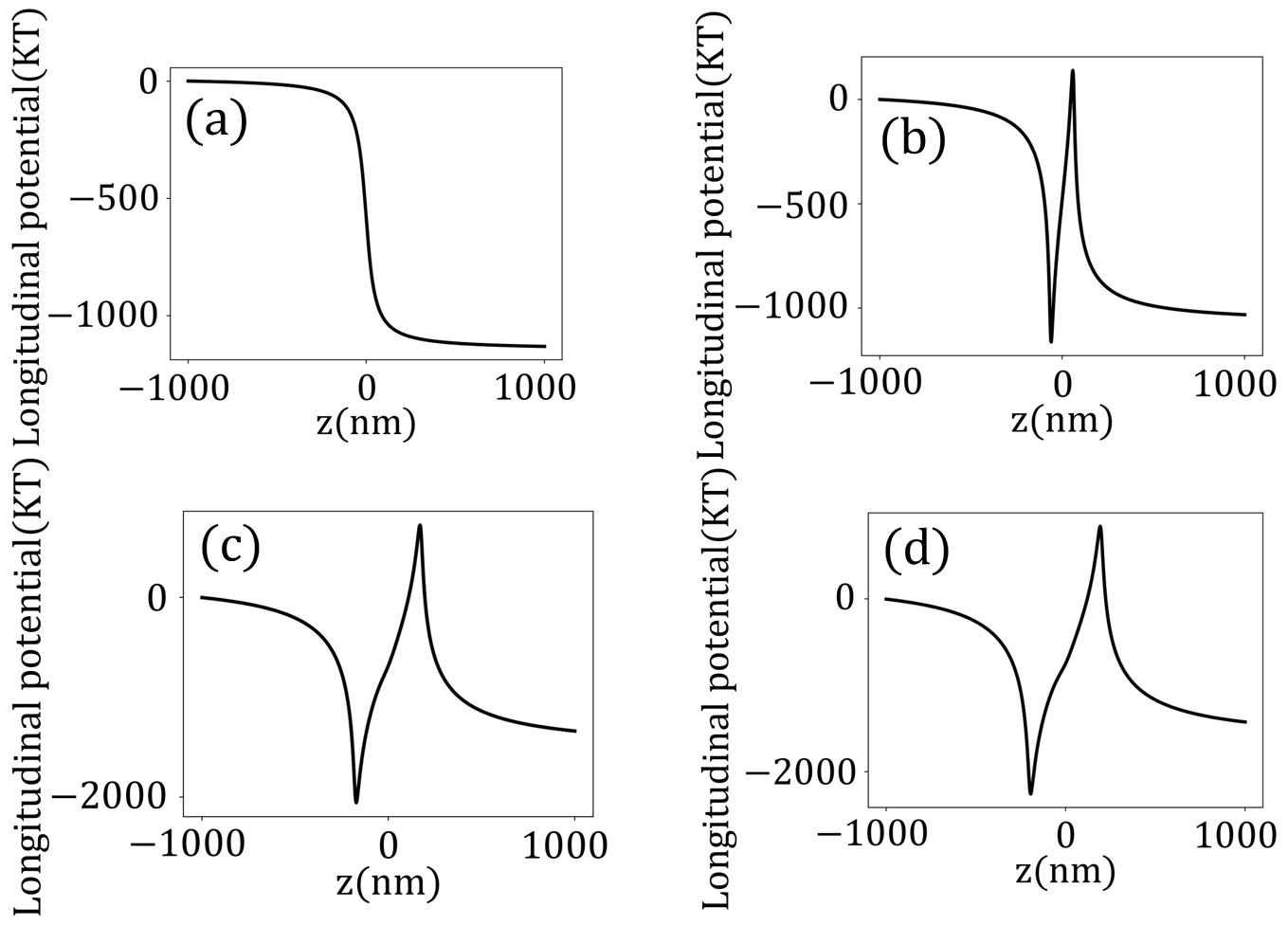}
\caption{Longitudinal potential for (a) $P_{ave}= 150 mW$, (b) $P_{ave}= 450 mW$, (c) $P_{ave}= 2000 mW$, and (d) $P_{ave}= 2500 mW$, taking into account both Kerr nonlinearity and two-photon absorption.}
\label{longitudinal TPA}
\end{figure}

\begin{figure}
\centering
\includegraphics[width=.48\textwidth]{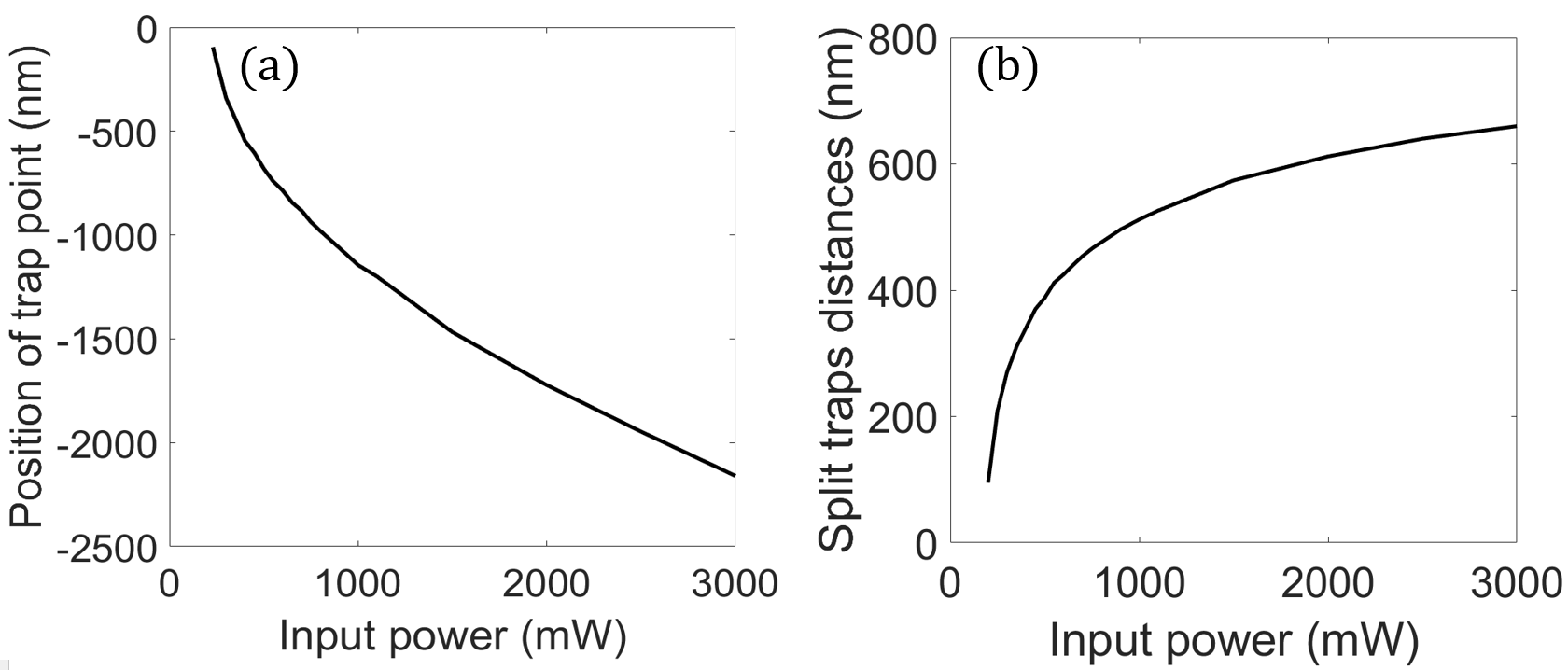}
\caption{(a) The spacing between two split traps as a function of  power in the transverse plane. (b) The positioning of trap points as a function of power in the longitudinal direction.}
\label{distance}
\end{figure}

 In the subsequent sections, we explore the effects of TPA and FWM on the optical trapping of gold nanoparticles.
 Figure \ref{longitudinal TPA} presents the longitudinal potential at different powers, taking into account a complex value for the third-order susceptibility, i.e., considering both Four-Wave Mixing (FWM) and TPA. Here, the complex susceptibility of gold is chosen to be $\chi_{3}=(3.9-6.6j)\times10^{-21} (m^2/V^2)$ \cite{de2008optical}. Additionally, the corresponding forces are detailed in the Supplemental Information. At low powers ($P_{ave}\le 150 mW$), gold nanoparticles exhibit instability in the longitudinal direction, as shown in Figure \ref{longitudinal TPA}(a). In this case, the effect of optical nonlinearities is not significant, and the trapping system behaves similarly to the linear regime. At higher powers ($P_{ave}> 150 mW$), an asymmetric potential well emerges, with the stable point situated to the left of the focal point, as depicted in Figures \ref{longitudinal TPA}(b)-(d). This stability trap occurs due to TPA, as we will demonstrate later. As the input power increases, the longitudinal potential energy maintains its asymmetric shape, and the position of the stable point progressively shifts to the left, as shown in Figure \ref{distance}(a). Moreover, by increasing power, the width of the potential trap increases. These behaviors in the longitudinal direction are predictions of our theory that have yet to be demonstrated experimentally.

\begin{figure}
\centering
\includegraphics[width=.48\textwidth]{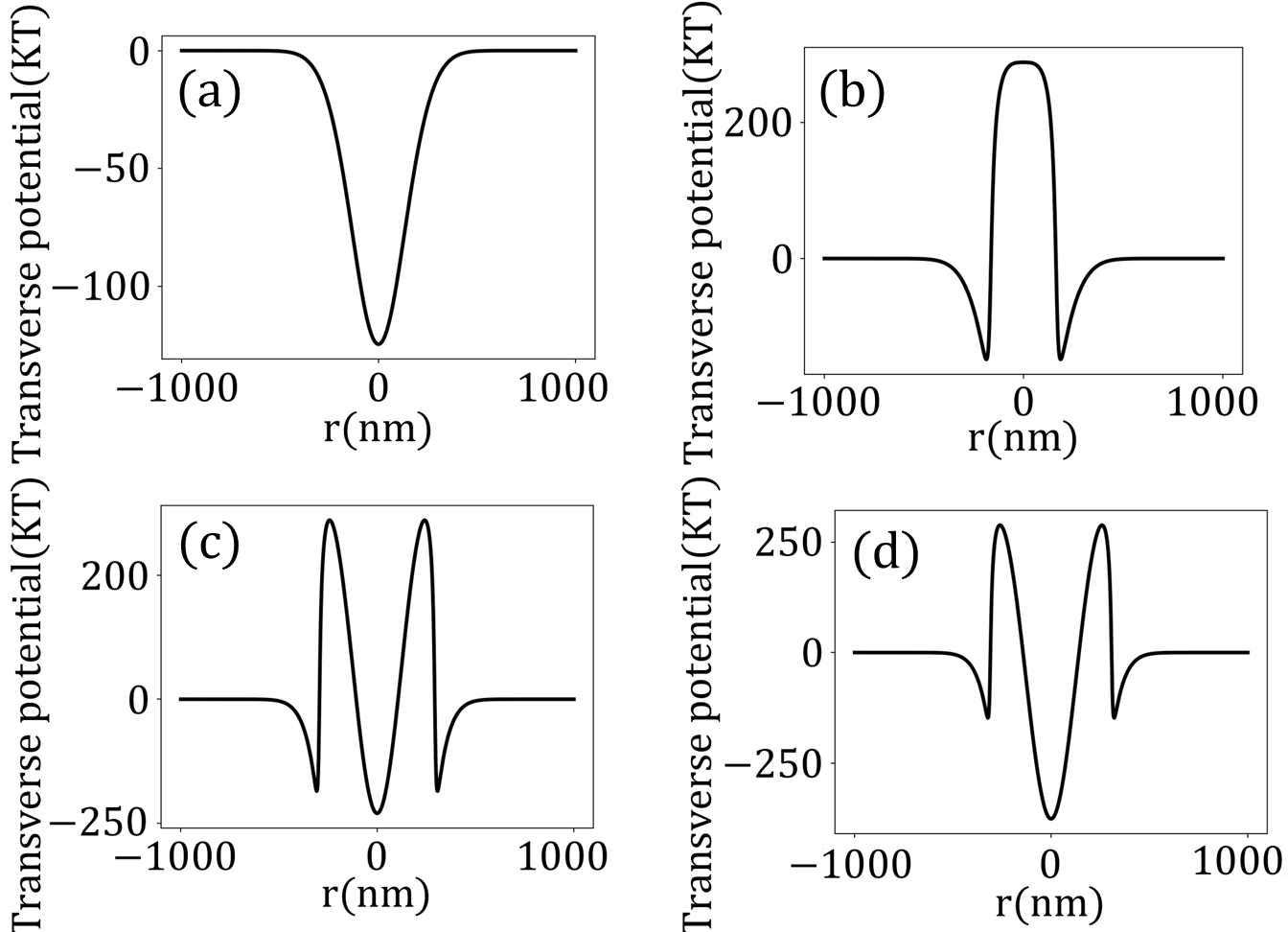}
\caption{Transverse potential energy for (a) $P_{ave}= 150 mW$, (b) $P_{ave}= 450 mW$, (c) $P_{ave}= 2000 mW$, and (d) $P_{ave}= 2500 mW$, accounting for two-photon absorption in conjunction with Kerr nonlinearity.}
\label{transverse TPA}
\end{figure}

 Figure \ref{transverse TPA} shows the transverse potential energies for different powers. Moreover, the corresponding forces are presented in the Supplemental Information. At lower powers ($P_{ave}\le 150 mW$), the potential energy resembles a quadratic shape, similar to the linear potential energy. However, at higher powers ($150 mW\le P_{ave}\le 1500 mW$), a trap split occurs with two wells symmetrically positioned on opposing sides of the focal point. The SA absorption likely happens in this regime which results in a bistable potential trap. The depths of these wells significantly exceed the kinetic energy of Brownian motion, ensuring stable trapping. Within the SA regime, an increase in power increases the potential barrier between the split traps, further impeding particle transitions between the traps.  At ultra-high powers ($P_{ave}>1500 mW$), an RSA regime is likely, characterized by a tri-stable potential trap with three wells: one central and two symmetrically positioned on either side. We established that the total potential trap can be viewed as a composite of linear and nonlinear potential traps. While the linear potential exhibits a single well at the origin, the nonlinear potential introduces two symmetrical off-center wells. In the SA regime, the potential barrier separating these off-center wells surpasses the depth of the linear potential. Conversely, in the RSA domain, the linear potential depth overcomes the potential barrier, engendering a third central well. Compared to the central potential well, the off-center wells have a narrower width, indicating that the stiffness of the nonlinear trap is greater than that of the linear one.
 Interestingly, the depth of the off-center wells remains constant regardless of input power (Figure \ref{transverse TPA} (b)-(c)), yet depends on TPA, as we will show explicitly later.

 The spacing between the two off-center split traps expands with increasing power as depicted in Figure \ref{distance} (b).  Notably, the variance in distance is more pronounced in the SA domain compared to its RSA counterpart. The distance between these off-center potential points can surpass the diffraction limit. For instance, at an input power of $P_{ave}= 200 mW$, this distance is $95nm$. These observations are in agreement with the experimental findings presented in \cite{jiang2010nonlinear}.

\begin{figure}
\centering
\includegraphics[width=.48\textwidth]{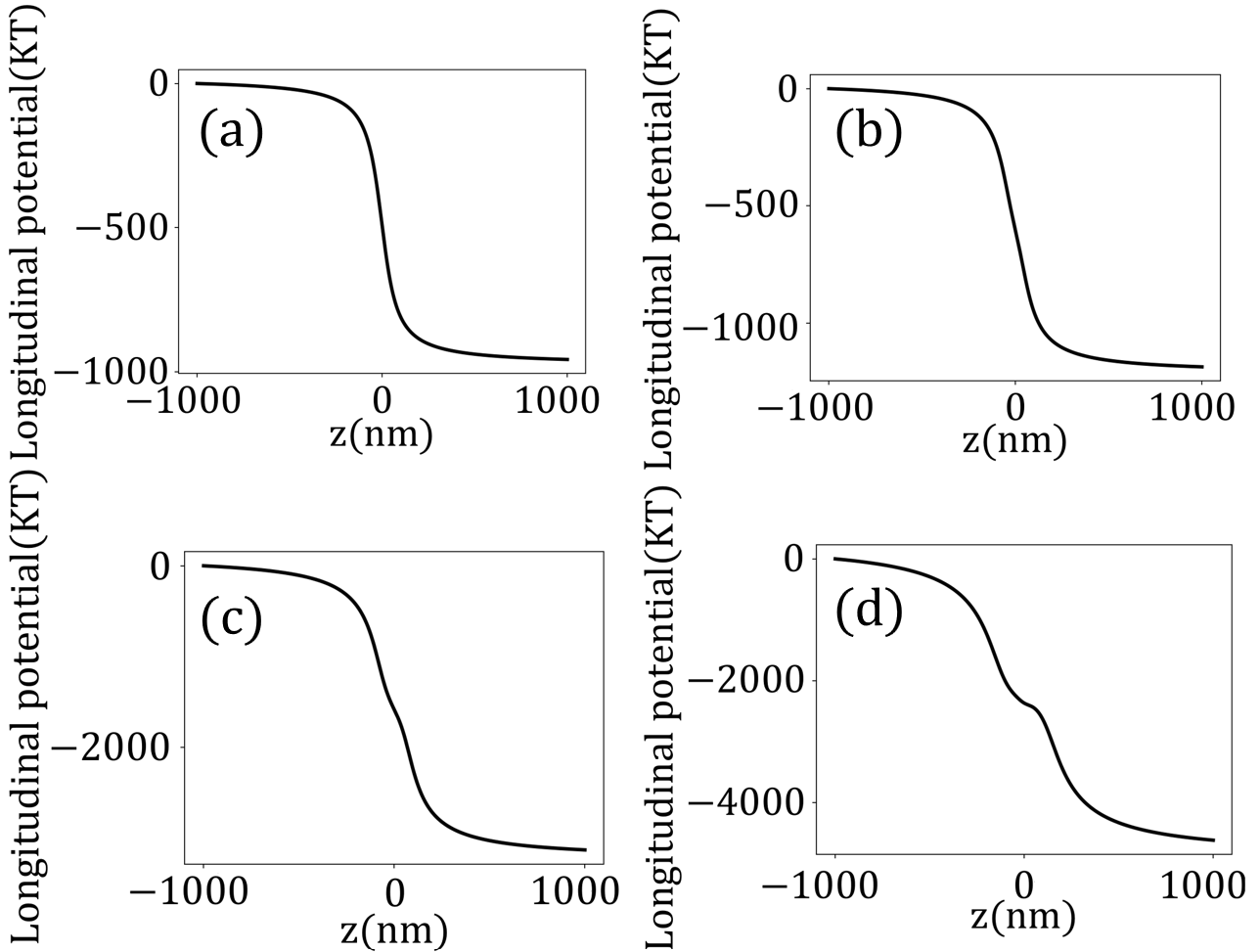}
\caption{Longitudinal potential energy for (a) $P_{ave}= 200 mW$, (b) $P_{ave}= 600 mW$, (c) $P_{ave}= 1100 mW$, and (d) $P_{ave}= 3000 mW$, when considering only FWM.}
\label{longitudinal FWM}
\end{figure}

In what follows, we only investigate the influence of FWM  on the optical trapping of gold nanoparticles. Figure \ref{longitudinal FWM} shows the longitudinal potential energies for various powers when considering only the FWM process, i.e., 
($\chi_3^{\prime\prime}=0$). Corresponding forces are again shown in the Supplemental Information. As shown, the gold nanoparticles are unstable in the longitudinal direction. Notably, by considering only FWM, gold nanoparticles remain unstable in the propagation direction. While the FWM introduces a slight perturbation to the longitudinal potential at high powers, the trap retains its instability.

\begin{figure}
\centering
\includegraphics[width=.48\textwidth]{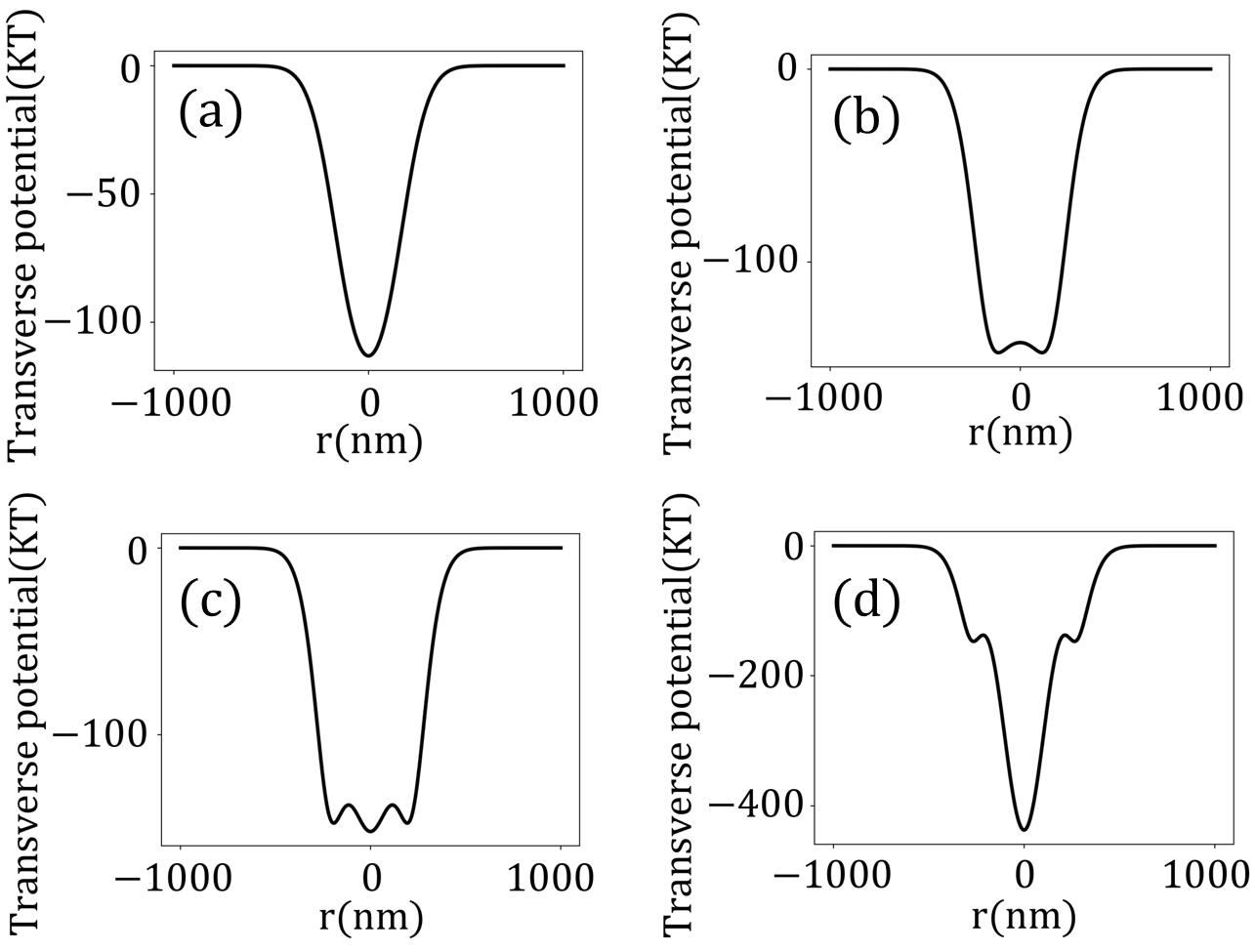}
\caption{Transverse potential energy for (a) $P_{ave}= 200 mW$, (b) $P_{ave}= 600 mW$, (c) $P_{ave}= 1100 mW$, and (d) $P_{ave}= 3000 mW$, when considering only the FWM process.}
\label{transverse FWM}
\end{figure}

Figure \ref{transverse FWM} shows the transverse potential energies, and, the corresponding forces are presented in the Supplemental Information.
At low powers ($P_{ave}<200 mW$), the FWM impact of nonlinearity is minimal, rendering the potential energy as a simple quadratic form with a singular well. Within medium powers ($200 mW\le P_{ave}\le 500 mW$), the potential energy splits into two shallow wells symmetrically positioned on either side of the focal point. At higher powers ($500 mW\le P_{ave}\le 2500 mW$), the potential energy splits into three shallow wells: a central one at the focal point and two equidistant ones around the origin. In ultra-high powers ($P_{ave}\ge2500 mW$), the central well deepens, while the depths of the off-central wells remain unaltered. These localized potential wells are superficial, preventing particles from remaining confined. Once trapped in one well, Brownian motion easily moves particles to transition between wells.

 Figures \ref{longitudinal FWM} and \ref{transverse FWM} reveal that the FWM process has little contribution to forming stable split traps observed experimentally. Even though it introduces shallow, split traps in the transverse plane, these traps lack stability. Moreover, it does not contribute to the longitudinal stability. These findings contradict experimental observations wherein particles are consistently trapped within split traps on the transverse plane or stably positioned longitudinally. These results show that the origin of split traps, as well as their longitudinal stability, is attributed to TPA.

The subsequent section provides a physical interpretation of the nonlinear trapping system. In the SA regime, the transverse potential traps exhibit behavior similar to the bi-stable potential of a non-harmonic oscillator. In contrast, within the RSA regime, the system behaves like a tri-stable potential. As previously demonstrated, the total trapping energy is a combination of linear and nonlinear potentials. Accordingly, the bi-stable and tri-stable potential energies can be presented by the equations $U_{bi}=-\frac{1}{2}k_{1}x^2+\frac{1}{4}k_{2}x^4$, and $U_{tri}=\frac{1}{2}k_{1}x^2-\frac{1}{4}k_{2}x^4+\frac{1}{6}k_{3}x^6$, respectively. Here, $k_{1}$ is linear stiffness and $k_{2}$ and $k_{3}$  denote nonlinear stiffnesses. To better understand the mechanism of nonlinear trapping, we make some simplifications. Specifically, we neglect the scattering force and self-induced back action, justifiable assumptions considering that gold nanoparticles trapped in water have a reduced scattering cross-section in the transverse plane.

After simplifications, the total potential can be expressed as a combination of linear (first part) and nonlinear (second part) counterparts (the details of the calculations are presented in the Supplemental Information).
\begin{equation}
    U = -E_0^2 e^{-2\rho^2} + \frac{3(3+\chi^{\prime})}{(3+\chi^{\prime})^2+(\chi^{\prime\prime})^2}E_0^2 e^{-2\rho^2}
    \label{equation 11}
\end{equation}

As shown, the nonlinear potential is enveloped by the linear one. The magnitude of total potential at the focal point, i.e. $U(\rho=0)$ reads  
\begin{equation}
    U(0) = -E_0^2+ \frac{3(3+\chi_1^{\prime} + 3\chi_3^{\prime}E_0^2)}{(3+\chi_1^{\prime} + 3\chi_3^{\prime}E_0^2)^2+(\chi_1^{\prime\prime} + 3\chi_3^{\prime\prime}E_0^2)^2}E_0^2
    \label{equation 12}
\end{equation}

Here, the first term represents the depth of the linear potential, and the second term corresponds to the amplitude of the nonlinear potential barrier, with $E_0$
being the maximum amplitude of the electric field. As the power increases, the depth of the linear potential deepens, while the potential barrier of the nonlinear potential initially rises (in the SA regime) before converging to a fixed value (in the RSA regime). In the SA regime, the magnitude of the second term dominates the first, leading to positive amplitudes as confirmed in Figure \ref{transverse TPA}(b). Conversely, in the RSA regime, the magnitude of the nonlinear term at the focal point remains unchanged with increasing power. In contrast, the depth of the linear potential continues to deepen. Thus, the magnitude of the linear component prevails over the nonlinear one, resulting in the emergence of a centered well, as depicted in Figures \ref{transverse TPA}(c) and (d). On the other hand, the off-center trap points happen at 
\begin{equation}
     \rho_{\pm} = \pm \sqrt{Ln(\frac{3(\chi_3^{\prime}-\chi_3^{\prime\prime})E_0^2}{\chi_1^{\prime\prime}-\chi_1^{\prime}-3})^{\frac{1}{2}}}
     \label{equation 13}
\end{equation}

The condition for forming a split trap is that the argument of the logarithmic function must be positive; i.e. $\frac{3(\chi_3^{\prime}-\chi_3^{\prime\prime})E_0^2}{\chi_1^{\prime\prime}-\chi_1^{\prime}-3}>1$. According to equation \ref{equation 13}, the distance between off-center points is proportional to $\sqrt{Ln({E_0})}$ a behavior illustrated in Figure \ref{distance}(a) through a more exact approach. Moreover, the depth of potential wells at off-centered wells donated $U(\rho_{\pm})$ is given by

\begin{equation}
    U(\rho_{\pm}) = \frac{1}{2}\frac{\chi_1^{\prime\prime}-(\chi_1^{\prime}+3)}{\chi_1^{\prime\prime}\chi_3^{\prime}-\chi_3^{\prime\prime}(3+\chi_1^{\prime})}
    \label{equation14}
\end{equation}

As shown, the depth of the off-center trap points is independent of the input power. Moreover, the first part of the denominator is much smaller than the second part, which indicates that the depth of the off-center traps is primarily determined by TPA. These findings agree with the findings in Figures \ref{transverse TPA} and \ref{transverse FWM}.

By comparing $U_{bi}$ and $U$, the coupling between linear and nonlinear stiffnesses can be calculated as follows:

\begin{equation}
    \frac{k_1}{k_2} = \rho_{+}=\ln(\frac{3(\chi_3^{\prime}-\chi_3^{\prime\prime})E_0^2}{\chi_1^{\prime\prime}-\chi_1^{\prime}-3})^{\frac{1}{2}}
    \label{equation15}
\end{equation}

This equation shows that the coupling between the two oscillators is not constant; instead, it varies depending on the power, as well as linear and nonlinear susceptibility parameters. Furthermore, the value of $\rho_{+}$ is on the order of micrometers, indicating that $k_2$ is significantly larger than  $k_1$, as previously demonstrated. Similar analytical methods can also be applied to a tri-stable potential.

\section{Conclusion} 

This paper presents a new theoretical approach to the nonlinear optical trapping of nanoparticles, emphasizing the pivotal roles of four-wave mixing and two-photon absorption. The research highlights TPA's key role in longitudinal trapping stability and reveals multiple split traps in diverse absorption settings. It suggests that the third-order nonlinearity of gold nanoparticles at plasmon resonances can explain the stable longitudinal trapping observed experimentally. By comparing the nonlinear trap system to a nonlinear harmonic oscillator, the paper deepens our understanding of dipole-regime trapping generally. Ultimately, this study fills existing gaps and advances the field of optical trapping, enriching insights into light-matter interactions.
\begin{acknowledgments}
We would like to thank Dr. Camacho for his useful discussions, as well as Spencer Duke for his contributions to this work.

% REV\TeX{}, offering suggestions and encouragement, testing new versions,
% \dots.
\end{acknowledgments}

\end{document}

%labels equations with S1, S2, etc.
\setcounter{equation}{0}
\renewcommand{\theequation}{S\arabic{equation}}

\renewcommand{\thefigure}{S\arabic{figure}}

\title{Nonlinear Optical Trapping of Gold Nanoparticles}

\preprint{prl/123-QED}

 \author{S. Mirzaei-Ghormish}
 \affiliation{Department of Electrical and Computer Engineering, Brigham Young University, Provo, UT}%Lines break automatically or can be forced with \\
 \author{K. Qaderi}%
 \affiliation{Department of Electrical and Computer Engineering, Brigham Young University, Provo, UT}
% \affiliation{ 
% Authors' institution and/or address%\\This line break forced with \textbackslash\textbackslash
% }%

 \author{D. Smalley}
\affiliation{Department of Electrical and Computer Engineering, Brigham Young University, Provo, UT}
 \email{smalley@byu.edu.}

% \author{S. Mirzaei-Ghormish}
 %\affiliation{Department of Electrical and Computer Engineering, Brigham Young University, Provo, UT}%Lines break automatically or can be forced with \\

%  \homepage{http://www.Second.institution.edu/~Charlie.Author.}
% \affiliation{%
% Second institution and/or address%\\This line break forced% with \\
% }%

%  \homepage{http://www.Second.institution.edu/~Charlie.Author.}
% \affiliation{%
% Second institution and/or address%\\This line break forced% with \\
% }%

\date{\today}% It is always \today, today,

\maketitle
\section{Theoretical calculations}
In this section, we develop the optical force calculations based on nonlinear polarization. In the dipole regime, the total force acting on the particle has two components: gradient and scattering force. These force components depend on real and imaginary parts of effective polarizability, respectively. When considering the nonlinear susceptibility, the static polarizability can be written as follows:
\begin{equation}
    \alpha_{0} = 4\pi\epsilon_oR^3\frac{\epsilon_{l}+3\chi_3E^2-\epsilon_{m}}{\epsilon_{l}+3\chi_3E^2+2\epsilon_{m}}
    \label{equation1}
\end{equation}
where $\epsilon_{l}$ is linear permittivity of the particle . 
 For simplicity in the calculations, we consider the medium to be air, thus:

\begin{equation}
    \alpha_{0} = 4\pi\epsilon_oR^3\frac{\epsilon_{l}+3\chi_3E^2-1}{\epsilon_{l}+3\chi_3E^2+2}= 4\pi\epsilon_oR^3\frac{(\epsilon_{l}-1)(1+\frac{3\chi_3E^2}{\epsilon_{l}-1})}{(\epsilon_{l}+2)(1+\frac{3\chi_3E^2}{\epsilon_{l}+2})}
    \label{equation2}
\end{equation}

In general, the nonlinear parts are much smaller than their linear counterparts, i.e. $|\frac{3\chi_3E^2}{\epsilon_{l}+2}|<1$. Thus, we can use the Taylor expansion of the denominator as follows: 
\begin{equation} 
    \alpha_{0} = 4\pi\epsilon_oR^3\frac{(\epsilon_{l}-1)}{(\epsilon_{l}+2)}(1+\frac{3\chi_3E^2}{\epsilon_{l}-1})(1-\frac{3\chi_3E^2}{\epsilon_{l}+2})=4\pi\epsilon_oR^3\frac{(\epsilon_{l}-1)}{(\epsilon_{l}+2)}(1+\frac{3\chi_3E^2}{\epsilon_{l}-1}-\frac{3\chi_3E^2}{\epsilon_{l}+2})
    \label{equation3}
\end{equation}

Moreover, $E^4$ terms are much smaller than the other terms. After removing these terms we get: 

\begin{equation}
    \alpha_{0} = 4\pi\epsilon_oR^3\frac{(\epsilon_{l}-1)}{(\epsilon_{l}+2)}+12\pi\epsilon_oR^3\chi_3E^2\frac{3}{(\epsilon_{l}+2)^2}
    \label{equation4}
\end{equation}

The first term is the linear part, and the second term is the nonlinear part of static susceptibility ($\alpha_{0}=\alpha^{(L)}_{0}+\alpha^{(NL)}_{0}$). The linear static polarizability in terms of susceptibilities is written as
\begin{equation}
    \alpha_{0}^{(L)} = 4\pi\epsilon_oR^3\frac{\chi^{\prime}+i\chi^{\prime\prime}}{\chi^{\prime}+i\chi^{\prime\prime}+3}= 4\pi\epsilon_oR^3\frac{(\chi^{\prime}+i\chi^{\prime\prime})(3+\chi^{\prime}-i\chi^{\prime\prime})}{(3+\chi^{\prime})^2+(\chi^{\prime\prime})^2}
    \label{equation5}
\end{equation}
And in term of real $\alpha_{0,r}^{(L)}$ and imaginary  $\alpha_{0,I}^{(L)}$ components  it can be written as
\begin{equation}
    \alpha_{0,r}^{(L)} = 4\pi\epsilon_oR^3\frac{\chi^{\prime}(3+\chi^{\prime})+(\chi^{\prime\prime})^2}{(3+\chi^{\prime})^2+(\chi^{\prime\prime})^2}
    \label{equation6}
\end{equation}

\begin{equation}
    \alpha_{0,I}^{(L)} = 12\pi\epsilon_oR^3\frac{\chi^{\prime\prime}}{(3+\chi^{\prime})^2+(\chi^{\prime\prime})^2}
    \label{equation7}
\end{equation}
On the other hand, the nonlinear static polarizability is 
\begin{equation}
    \alpha_{0}^{(NL)} = 12\pi\epsilon_oR^3E^2\frac{3\chi_3}{(\epsilon_{1}+2)^2} = 36\pi\epsilon_oR^3E^2\frac{\chi_3}{(\chi+3)^2}
    \label{equation8}
\end{equation}
where its real $ \alpha_{0,r}^{(NL)}$ and imaginary $ \alpha_{0,I}^{(NL)}$ components of nonlinear static polarizability are calculated as follows 
\begin{equation}
    \alpha_{0,r}^{(NL)} = \frac{36\pi\epsilon_oR^3E^2}{[(3+\chi^\prime)^2+(\chi^{\prime\prime})^2]^2}[(3+\chi^\prime )^2\chi_3^{\prime} + 2(3+\chi^\prime)\chi^{\prime\prime}\chi_3^{\prime\prime} - \chi_3^{\prime}(\chi^{\prime\prime})^2]
    \label{equation9}
\end{equation}

\begin{equation}
    \alpha_{0,I}^{(NL)} = \frac{36\pi\epsilon_oR^3E^2}{[(3+\chi^\prime)^2+(\chi^{\prime\prime})^2]^2}[(3+\chi^\prime)^2\chi_3^{\prime\prime} - 2(3+\chi^\prime)\chi^{\prime\prime}\chi_3^{\prime} - \chi_3^{\prime\prime}(\chi^{\prime\prime})^2]
    \label{equation10}
\end{equation}

 By substituting the static polarizability into effective polarizability  $\alpha_{0}/(1-i\frac{\alpha_{0}k^3}{6\pi\epsilon_{0}})$ we get:
\begin{equation}
    \alpha = \frac{\alpha_{0}^{(L)} + \alpha_{0}^{(NL)}}{1-i\frac{k^3}{6\pi\epsilon_{0}}(\alpha_{0}^{(L)} + \alpha_{0}^{(NL)})}= \frac{(\alpha_{0,r}^{(L)}+\alpha_{0,r}^{(NL)}) + i(\alpha_{0,I}^{(L)}+\alpha_{0,I}^{(NL)})}{1-i\frac{k^3}{6\pi\epsilon_{o}}\left[ (\alpha_{0,r}^{(L)}+\alpha_{0,r}^{(NL)}) + i(\alpha_{0,I}^{(L)}+\alpha_{0,I}^{(NL)})  \right]}
     \label{equation11}
\end{equation}

\begin{equation}
    \alpha = \frac{\left[  (\alpha_{0,r}^{(L)}+\alpha_{0,r}^{(NL)}) + i(\alpha_{0,I}^{(L)}+\alpha_{0,I}^{(NL)})\right] \left[ 1+\frac{k^3}{6\pi\epsilon_{o}}(\alpha_{0,I}^{(L)}+\alpha_{0,I}^{(NL)})+i\frac{k^3}{6\pi\epsilon_{o}}(\alpha_{0,r}^{(L)}+\alpha_{0,r}^{(NL)}) \right]}{\left[ 1+\frac{k^3}{6\pi\epsilon_{o}}(\alpha_{0,I}^{(L)}+\alpha_{0,I}^{(NL)})\right]^2+\left[ \frac{k^3}{6\pi\epsilon_{o}}(\alpha_{0,r}^{(L)}+\alpha_{0,r}^{(NL)}) \right]^2}
    \label{equation12}
\end{equation}
The terms associated with $(\alpha_{0,r}^{(NL)})^2$ and $(\alpha_{0,I}^{(NL)})^2$ are much smaller than other parts; therefore, we remove them. After simplifications, the real and imaginary effective polarizabilities read

\begin{equation}
    \alpha^{\prime} = \frac{\alpha_{0,r}^{(L)}+\alpha_{0,r}^{(NL)}}{\left[ 1+\frac{k^3}{6\pi\epsilon_{0}}(\alpha_{0,I}^{(L)}+\alpha_{0,I}^{(NL)})\right]^2+\left[ \frac{k^3}{6\pi\epsilon_{0}}(\alpha_{0,r}^{(L)}+\alpha_{0,r}^{(NL)}) \right]^2}
    \label{equation13}
\end{equation}

\begin{equation}
    \alpha^{\prime\prime} = \frac{(\alpha_{0,I}^{(L)}+\alpha_{0,I}^{(NL)})+\frac{k^3}{6\pi\epsilon_{o}}\left[ (\alpha_{0,r}^{(L)}+\alpha_{0,r}^{(NL)})^2 + (\alpha_{0,I}^{(L)}+\alpha_{oI}^{(NL)})^2 \right]
    }{\left[ 1+\frac{k^3}{6\pi\epsilon_{o}}(\alpha_{0,I}^{(L)}+\alpha_{0,I}^{(NL)})\right]^2+\left[ \frac{k^3}{6\pi\epsilon_{o}}(\alpha_{0,r}^{(L)}+\alpha_{0,r}^{(NL)}) \right]^2}
    \label{equation14}
\end{equation}
These equations can be more simplified by eliminating smaller terms as follows:
\begin{equation}
    \alpha^{\prime} = \alpha_{o,r}^{(L)}+\alpha_{o,r}^{(NL)}
     \label{equation15}
\end{equation}

\begin{equation}
    \alpha^{\prime\prime} = (\alpha_{0,I}^{(L)}+\alpha_{0,I}^{(NL)})+\frac{k^3}{6\pi\epsilon_{o}}\left[ (\alpha_{0,r}^{(L)})^2+(\alpha_{0,I}^{(L)})^2+2(\alpha_{0,r}^{(L)}\alpha_{0,r}^{(NL)} + \alpha_{0,I}^{(L)}\alpha_{0,I}^{(NL)}) \right]
     \label{equation16}
\end{equation}

By inserting equations \ref{equation6},\ref{equation7},\ref{equation9}, and \ref{equation10} into \ref{equation15} and \ref{equation16}, we can get the  final equations for effective polarizability. 

Our formulation can be applied to any type of particle including, non-absorbing, weakly absorbing, and highly absorbing particles. In these cases, the above equations can be written in more compact forms. For non-absorbing particles ($\chi^{\prime\prime}_1=0, \chi^{\prime\prime}_3=0$), they can be simplified as follows:
\begin{equation}
    \alpha_1^{\prime} = 4\pi\epsilon_oR^3\frac{\epsilon^{\prime}_{p}-1}{\epsilon^{\prime}_{p}+2} + 36\pi\epsilon_oR^3E^2\chi_3^{\prime}\frac{\epsilon^{\prime}_{p}-1}{(\epsilon^{\prime}_{p}+2)^2}
    \label{equation17}
\end{equation}

\begin{equation}
    \alpha_1^{\prime\prime} = \frac{8}{3}\pi\epsilon_ok^3R^6(\frac{\epsilon^{\prime}_{p}-1}{\epsilon^{\prime}_{p}+2})^2 + 48\pi\epsilon_ok^3R^6E^2\chi_3^{\prime}\frac{\epsilon^{\prime}_{p}-1}{(\epsilon^{\prime}_{p}+2)^3}
    \label{equation18}
\end{equation}

As demonstrated, FWM influences both the gradient and scattering forces involved in optical trapping. However, its impact on the gradient force is more pronounced leading to increasing trapping stability. This observation is consistent with the results reported in \cite{devi2020generalized, mirzaei-ghormish_nonlinear_2023}.

In low absorption limit ($\chi^{\prime\prime} \ll \chi^\prime$, $\chi_3^{\prime\prime}\ll \chi_3^\prime$), we get
\begin{equation}
    \alpha^{\prime} = \alpha_1^{\prime}
    \label{equation19}
\end{equation}
\begin{equation}
    \alpha^{\prime\prime} = \alpha_1^{\prime\prime}+12\pi\epsilon_0R^3\frac{\chi^{\prime\prime}+3E^2\chi_3^{\prime\prime}}{(\epsilon^{\prime}_{p}+2)^2}
    \label{equation20}
\end{equation}

In this condition, the FWM contributes to both gradient and scattering forces. Conversely, TPA solely contributes to the scattering force. The ratio of the FWM to TPA components in equation \ref{equation20} is proportional to $(kR)^3\left(\frac{\chi^{\prime}_3}{\chi^{\prime\prime}_3}\right)$, where $(kR)^3 \ll 1$ and $\left(\frac{\chi^{\prime}_3}{\chi^{\prime\prime}_3}\right) \gg 1$. As a result, in certain circumstances, the effect of FWM dominates, whereas in others, TPA primarily contributes to the scattering force. Depending on the values of $\chi^{\prime}_{3}$ and $\chi^{\prime\prime}_{3}$, as well as the magnitude of the input power, the FWM and TPA effects can either increase or decrease the gradient and scattering forces. Hence, the new potential traps other than the harmonic quadratic potential can be achieved.

\begin{figure}
    \centering
     \includegraphics[width=.6\textwidth]{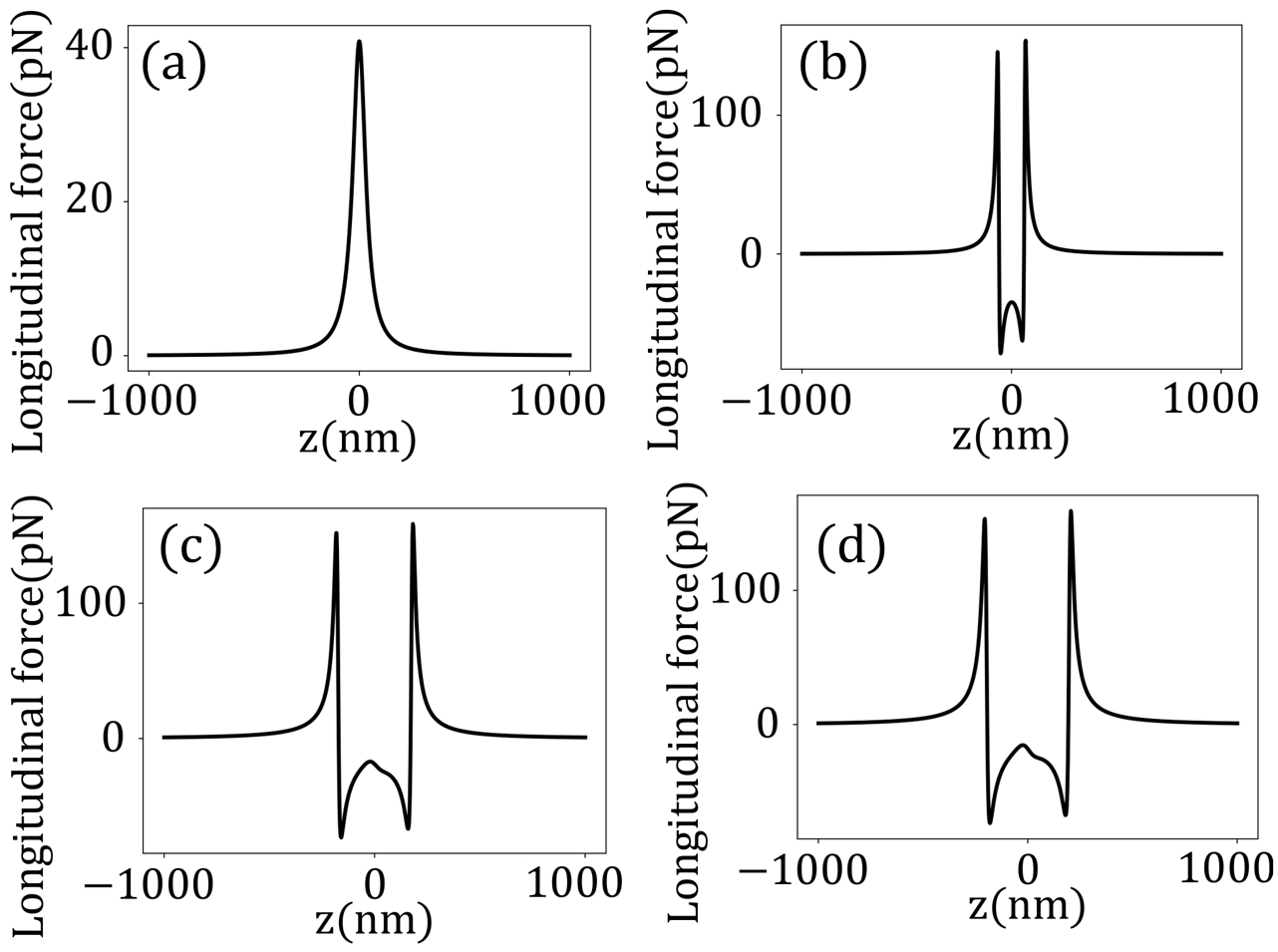}
     \caption{The longitudinal force for (a) $P_{ave}= 150 mW$, (b)  $P_{ave}= 450 mW$  (c) $P_{ave}= 2000 mW$, and (d) $P_{ave}= 2500 mW$, when considering both four-wave mixing and two-photon absorption} 
    \label{longitudinal TPA force}
\end{figure}

Figure \ref{longitudinal TPA force} shows the longitudinal forces at different powers when considering four-wave mixing and two-photon absorption. At lower average powers, no stable trap is created. As the power increases, two zero-force points with high stiffnesses appear; the left zero-force point is the stable point, and the second point is the unstable point. Further increasing the power causes the distance between these zero-force points to increase. At extremely high powers, another zero-point force tends to appear in the center. However, it does not lead to a stable trap.

\begin{figure}
    \centering
     \includegraphics[width=.6\textwidth]{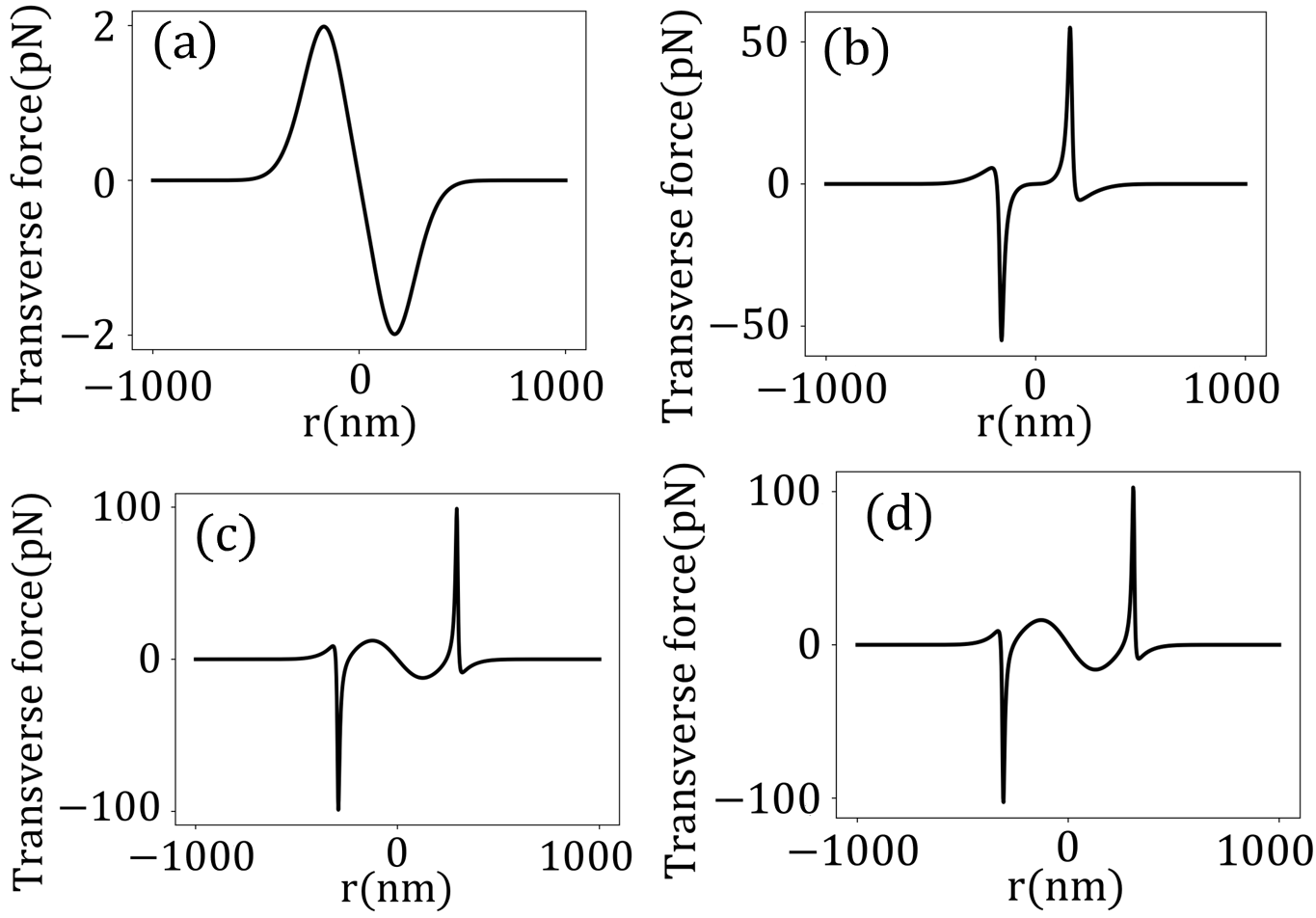}
     \caption{The transverse force for (a) $P_{ave}= 150 mW$, (b)  $P_{ave}= 450 mW$  (c) $P_{ave}= 2000 mW$, and (d) $P_{ave}= 2500 mW$, when considering both four-wave mixing and two-photon absorption.} 
    \label{transverse TPA force}
\end{figure}

Figure \ref{transverse TPA force} shows the transverse forces at varying power levels when accounting for both four-wave mixing and two-photon absorption. At low average powers, the transverse force resembles the linear force. As the power increases, two zero-force points emerge on the left and right sides of the center, corresponding to the two split trapping wells (nonlinear traps). At higher powers, three zero-force points become apparent: one is in the center (linear trap), and the other two are located on the left and right sides of the center, respectively. At extremely high powers, the stiffness of the central trap is enhanced, while the stiffness of the side traps remains unchanged. Moreover, the stiffness of nonlinear traps is much stronger than that of the central linear trap. 

\begin{figure}
    \centering
     \includegraphics[width=.6\textwidth]{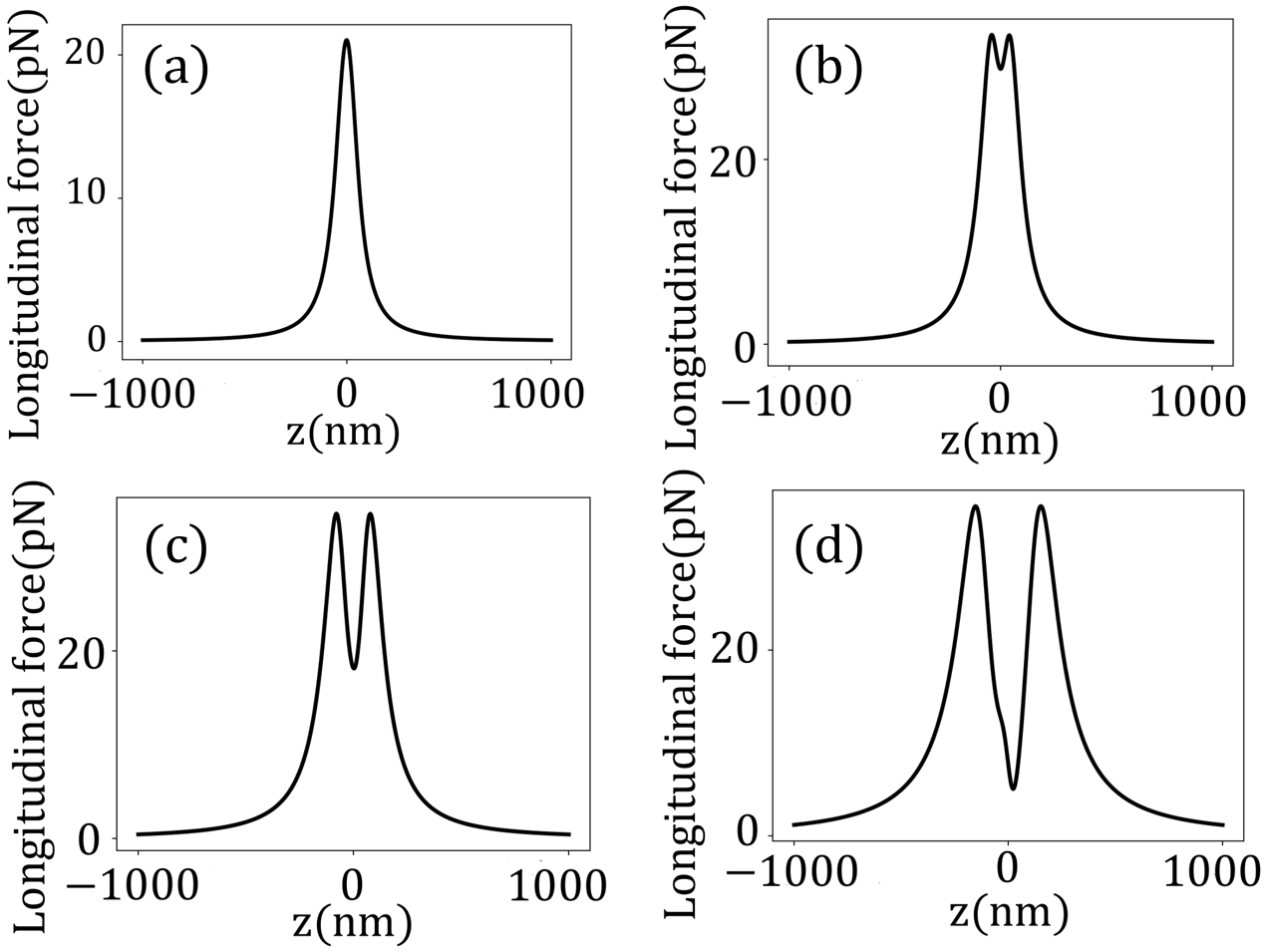}
     \caption{The longitudinal force for (a) $P_{ave}= 200 mW$, (b)  $P_{ave}= 600 mW$  (c) $P_{ave}= 1100 mW$, and (d) $P_{ave}= 3000 mW$, when considering four-wave mixing} 
    \label{longitudinal Kerr foce}
\end{figure}

Figure \ref{longitudinal Kerr foce} shows the longitudinal forces at different powers when considering only a four-wave mixing process. As the power increases, the longitudinal force splits into two branches; however, no zero-force point occurs along the axial direction. Therefore, four-wave mixing does not lead to longitudinal trap stability.

\begin{figure}
    \centering
     \includegraphics[width=.6\textwidth]{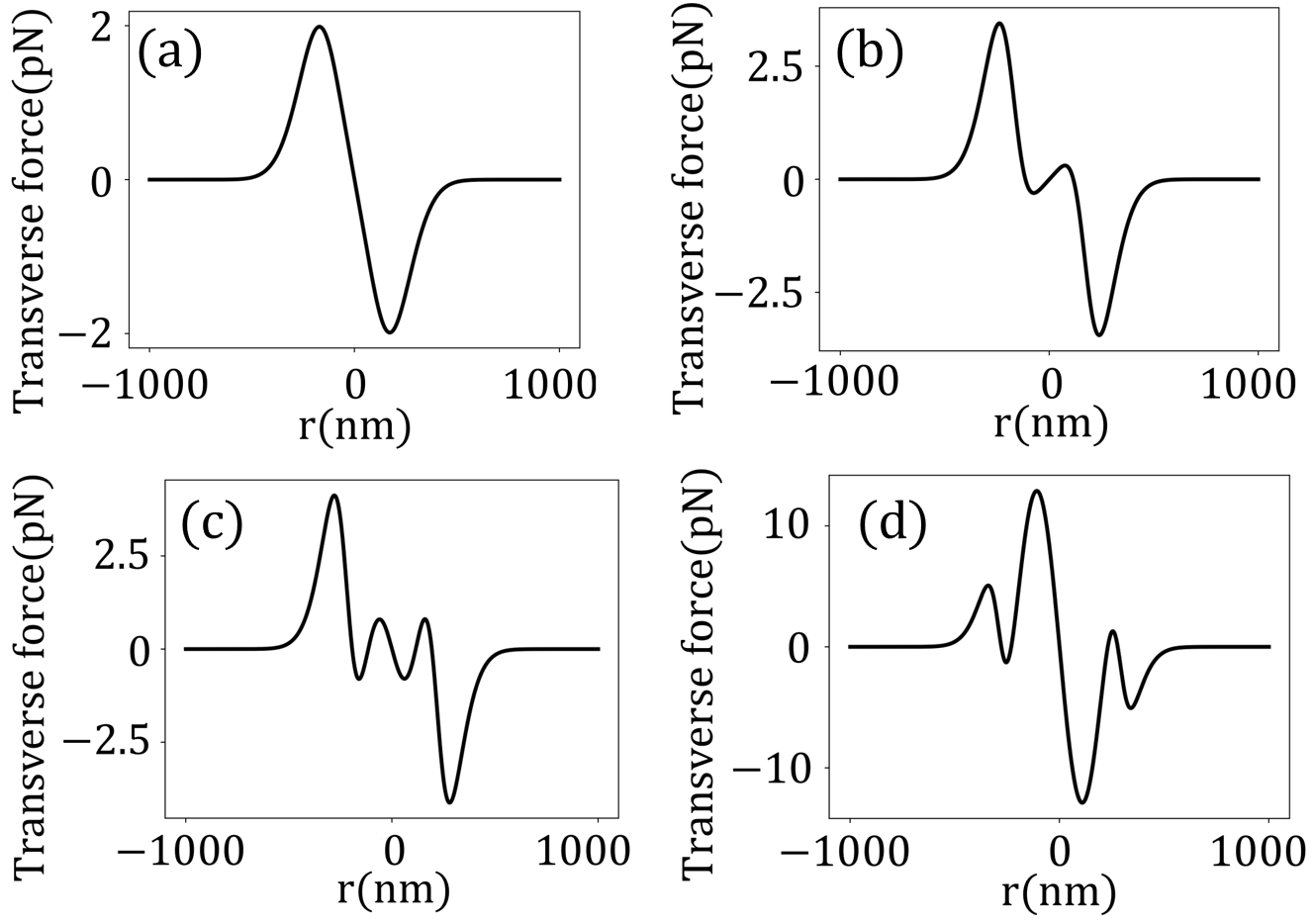}
     \caption{The transverse force for (a) $P_{ave}= 200 mW$, (b)  $P_{ave}= 600 mW$  (c) $P_{ave}= 1100 mW$, and (d) $P_{ave}= 3000 mW$, when considering four-wave mixing} 
    \label{transverse Kerr force}
\end{figure}

Figure \ref{transverse Kerr force} shows the transverse forces at different powers when considering only four-wave mixing. As shown, at low average powers, the transverse force resembles the linear force. With an increase in power, two zero-force points emerge on the left and right sides of the focal point, which correspond to the two-split trap wells. At high powers, three zero-force points become evident: one is at the center, and the other two are located on the left and right sides of the center. These correspond to three-split traps. At extreme powers, the stiffness of the central trap increases, while the stiffness of the side traps remains unchanged. Moreover, the stiffnesses of linear and nonlinear trap points in contrast to \ref{transverse TPA force} are almost constant. It should be mentioned that the linear and nonlinear stiffnesses are not sufficient to achieve a stable trap.

% \appendix

 \section{Physical interpretation}

In this section, we provide a physical interpretation of the nonlinear trapping system. In the transverse plane, we model the nonlinear system within the saturable absorption regime using a bistable potential and illustrate the linear and nonlinear stiffnesses in terms of the linear and nonlinear susceptibilities. A similar approach can be used to model the potential trap within a reverse saturable absorption regime using a tri-stable potential well.
 The nonlinear potential at distances between off-center points can be expressed by $U_{bi}  = -\frac{1}{2}k_1 x^2+\frac{1}{4}k_2 x^4$, where $k_1$ and $k_2$ respectively are the linear and nonlinear stiffnesses. The minimum points of this potential ($x = \pm \sqrt{\frac{k_1}{k_2}}$) are equivalent to the off-center trap points. To have a better insight into the mechanism of nonlinear optical trapping, we make some further simplifications. Here, we ignore the scattering force and self-induced back-action. These are reasonable assumptions because the gold nanoparticles experience a smaller scattering force in the transverse plane when they are immersed in water. In this case, the gradient potential reads $U = -\Re({\frac{\chi}{\chi + 3}})|\vec{E}|^2$ where 
$\vec{E} = \vec{E_0} e^{-\rho^2}$ is the Gaussian electric field with maximum amplitude of $E_0$, and total susceptibility is $\chi = \chi_1 + 3\chi_3 E_0^2 e^{-2\rho^2}$. Moreover, the real and imaginary parts of total susceptibilities can be written as $ \chi^{\prime} = \chi_1^{\prime} + \Delta \epsilon_3^{\prime} e^{-2\rho^2}$ and $\chi^{\prime\prime} = \chi_1^{\prime\prime} + \Delta \epsilon_3^{\prime\prime} e^{-2\rho^2} $ in which $\Delta \epsilon_3^{\prime} = 3\chi_3^{\prime}E_0^2$ and $\Delta \epsilon_3^{\prime\prime} = 3\chi_3^{\prime\prime}E_0^2$.

After simplifications, the potential can be expressed by:

\begin{equation}
    U = -E_0^2 e^{-2\rho^2} + \frac{3(3+\chi^{\prime})}{(3+\chi^{\prime})^2+(\chi^{\prime\prime})^2}E_0^2 e^{-2\rho^2}
    \label{equation 21}
\end{equation}

The total potential is expressed by the combination of linear (first term) and nonlinear (second term) potentials, i.e. $ U = U_L + U_{NL}$. The linear part has a minimum point at $\rho = 0$ where its depth is $U_L(0) = -E_0^2$. On the other hand, the nonlinear potential reads

\begin{equation}
    U_{NL} = \frac{3(3+\chi_1^{\prime} + \Delta \epsilon_3^{\prime} e^{-2\rho^2})}{(3+\chi_1^{\prime} + \Delta \epsilon_3^{\prime} e^{-2\rho^2})^2+(\chi_1^{\prime\prime} + \Delta \epsilon_3^{\prime\prime} e^{-2\rho^2})^2}E_0^2 e^{-2\rho^2}
    \label{equation 22}
\end{equation}

Its maximum happens when  \( e^{-2\rho^2} \) is eliminated which means it happens at \( \rho = 0 \) with the maximum potential of 
\begin{equation}
    U_{NL}(0) = \frac{3(3+\chi_1^{\prime} + 3\chi_3^{\prime}E_0^2)}{(3+\chi_1^{\prime} + 3\chi_3^{\prime}E_0^2)^2+(\chi_1^{\prime\prime} + 3\chi_3^{\prime\prime}E_0^2)^2}E_0^2.
    \label{equation 23}
\end{equation}

On the other hand, the minimum happens when $3+\chi_1^{\prime} + 3\chi_3^{\prime}e^{-2\rho^2} = \pm (\chi_1^{\prime\prime} + 3\chi_3^{\prime\prime}e^{-2\rho^2})$. When considering  $3+\chi_1^{\prime} + 3\chi_3^{\prime}e^{-2\rho^2} =  (\chi_1^{\prime\prime} + 3\chi_3^{\prime\prime}e^{-2\rho^2})$, the extremum points happen at $\rho^{2} =  {\ln(\frac{3(\chi_3^{\prime}-\chi_3^{\prime\prime})E_0^2}{\chi_1^{\prime\prime}-\chi_1^{\prime}-3})^{\frac{1}{2}}}$
and when considering $3+\chi_1^{\prime} + 3\chi_3^{\prime}e^{-2\rho^2} = - (\chi_1^{\prime\prime} + 3\chi_3^{\prime\prime}e^{-2\rho^2})$, the extremum points happen at $\rho^{2} ={\ln(\frac{\chi_1^{\prime\prime}+\chi_1^{\prime}+3}{3(\chi_3^{\prime}+\chi_3^{\prime\prime})E_0^2})^{\frac{1}{2}}}$. When the arguments of the logarithmic functions are larger than 1, we have the correct solutions and the off-center wells can appear. Therefore, whether split traps happen or not depends on the values of the linear and nonlinear susceptibilities and the input power. In our case, for ($\chi^{\prime}_3>0$,$\chi^{\prime\prime}_3<0$, $\chi^{\prime}_1<0$, and $\chi^{\prime\prime}_1>0$), the first extremum point is correct. Thus, the off-center trap points happen at 
\begin{equation}
     \rho_{\pm} = \pm \sqrt{\ln(\frac{3(\chi_3^{\prime}-\chi_3^{\prime\prime})E_0^2}{\chi_1^{\prime\prime}-\chi_1^{\prime}-3})^{\frac{1}{2}}}
     \label{equation 24}
\end{equation}
And the condition for happening split traps is: $\frac{3(\chi_3^{\prime}-\chi_3^{\prime\prime})E_0^2}{\chi_1^{\prime\prime}-\chi_1^{\prime}-3}>1$. 
Finally, the depth of nonlinear potential at off-center points is equal to

\begin{equation}
    U_{NL}(\rho_{\pm}) = \frac{1}{2}\frac{\chi_1^{\prime\prime}-(\chi_1^{\prime}+3)}{\chi_1^{\prime\prime}\chi_3^{\prime}-\chi_3^{\prime\prime}(3+\chi_1^{'})}
     \label{equation 25}
\end{equation}

 Then, by comparing the bistable potential ($U_{bi}$) with the trap potential ($U$) we can find the coupling between linear and nonlinear stiffnesses as follows:

\begin{equation}
    \frac{k_{1}}{k_{2}} = \ln(\frac{3(\chi_3^{\prime}-\chi_3^{\prime\prime})E_0^2}{\chi_1^{\prime\prime}-\chi_1^{\prime}-3})^\frac{1}{2}
    \label{equation 26}
\end{equation}